\documentstyle[11pt,epsfig,paspconf]{article}
\def\be{\begin{equation}}
\def\ee{\end{equation}}
\def\bea{\begin{eqnarray}}
\def\eea{\end{eqnarray}}

\def\etal{{\em et al.}            }

\def\kms{{\rm \,km\,s}^{-1}}

\def\hMpc{\,h^{-1}{\rm Mpc}}

\def\spose#1{\hbox to 0pt{#1\hss}}
\def\simlt{\mathrel{\spose{\lower 3pt\hbox{$\mathchar"218$}}
     \raise 2.0pt\hbox{$\mathchar"13C$}}}
\def\simgt{\mathrel{\spose{\lower 3pt\hbox{$\mathchar"218$}}
     \raise 2.0pt\hbox{$\mathchar"13E$}}}
\def\({\left(}
\def\){\right)}
\def\[{\left[}
\def\]{\right]}
\def\<{\left\langle}
\def\>{\right\rangle}

\def\AJ{{\em Astron.~J.~}}
\def\ApJ{{\em Astrophys.~J.~}}

\def\ApJL{{\em Astrophys.~J.\ Lett.~}}

\def\IAU130{in {\em Large Scale Structures of the Universe}, IAU Symposium 130~}
\def\MN{{\em Mon.Not.R.astr.Soc.~}}

\def\edcomment#1{\iffalse\marginpar{\raggedright\sl#1\/}\else\relax\fi}
\reversemarginpar

\begin{document}
\title{Density and Velocity Fields from the PSCz Survey}

\author{Will Saunders$^1$, Enzo Branchini$^2$, Luis Teodoro$^3$, Alan Heavens$^1$, Andy Taylor$^1$, Helen Valentine$^1$, Kenton D'Mellow$^1$,
Seb Oliver$^4$, Oliver Keeble$^4$, Michael Rowan-Robinson$^4$, Jacob Sharpe$^4$, Steve Maddox$^5$, Richard McMahon$^5$, George Efstathiou$^5$, 
Will Sutherland$^6$, Helen Tadros$^6$,  Bill Ballinger$^6$, Inga Schmolt$^6$,
Carlos Frenk$^7$, Simon White$^8$}
\affil{$^1$ Institute for Astronomy, University of Edinburgh;
$^2$  Kapteyn Institute, University of Groningen;
$^3$ Istituto Superior Tecnico, Lisboa; 
$^4$  Blackett Laboratory, Imperial College, University of London;
$^5$ Institute of Astronomy, Cambridge University;
$^6$  Dept. of Physics, Oxford University;
$^7$ Dept of Physics, University of Durham; \\
$^8$ MPI-Astrophysik, Garching}

\begin{abstract}

We present the results for the predicted density and peculiar velocity fields 
and the dipole from the PSCz survey of 15,000 IRAS galaxies over 84\%
of the sky. We find a significant component to the dipole arising between 
$6000$ and $15,000 \kms$, but no significant component from greater distances.  
The misalignment with the CMB is $20 \deg$. The most remarkable feature of the
PSCz model velocity field is a coherent large-scale flow along the 
baseline connecting Perseus-Pisces, the Local Supercluster, Great Attractor 
and the Shapley Concentration. 
We have measured the parameter $\beta$ 
using the amplitude of the dipole, bulk flow
and point by point comparisons between the
individual velocities of galaxies in the MarkIII and SFI datasets, and the large-scale clustering distortion in redshift space.
All our results are consistent with $\beta = 0.6 \pm 0.1$.

\end{abstract}
\section{Introduction}
\vspace{-5pt}

One of the motivations for the PSCz survey at its inception in 1992 was the 
huge effort going into peculiar velocity surveys; neither the QDOT nor the 1.2Jy 
surveys were deep and dense enough to provide a satisfactory model for 
the gravity field.
Our goals were to (a) maximise sky coverage, and (b) to obtain the best possible 
completeness and flux uniformity within well-defined area and redshift ranges. 
The survey consists of 15,000 IRAS galaxies and its sky coverage is 84\%. 
The median depth is just $8100 \kms$, although  useful information 
is available out to 
$30,000\kms$ at high latitudes and $15,000\kms$ everywhere.
A more detailed description of the survey specification
is given in Saunders \etal (1999). The distribution of identified galaxies and the mask are shown 
in figure 1. The $N(z)$ distribution is shown in figure 2a. 
The selection function $\psi(r)$ (defined here as the expected density of galaxies seen in 
the survey as a function of distance in the absence of clustering)
is derived using the methods of Mann \etal (1996) and shown in figure 2b.
The uncertainties are less than 10\% 
in the range $10-300\hMpc$. 

The knowledge of the selection function enables one to weigh galaxy properly 
and to  construct the (redshift-space) density field. A 3D view of this
is given in figure 3.
Figure 4 shows the PSCz density field in the Supergalactic Plane
after removing the effect of redshift space distortions.
A variable smoothing length increasing linearly along the radial direction
has been used. The continuous line shows the $\delta=0$ contour (Branchini \etal 1999).

\begin{figure}
\centerline{\epsfig{figure=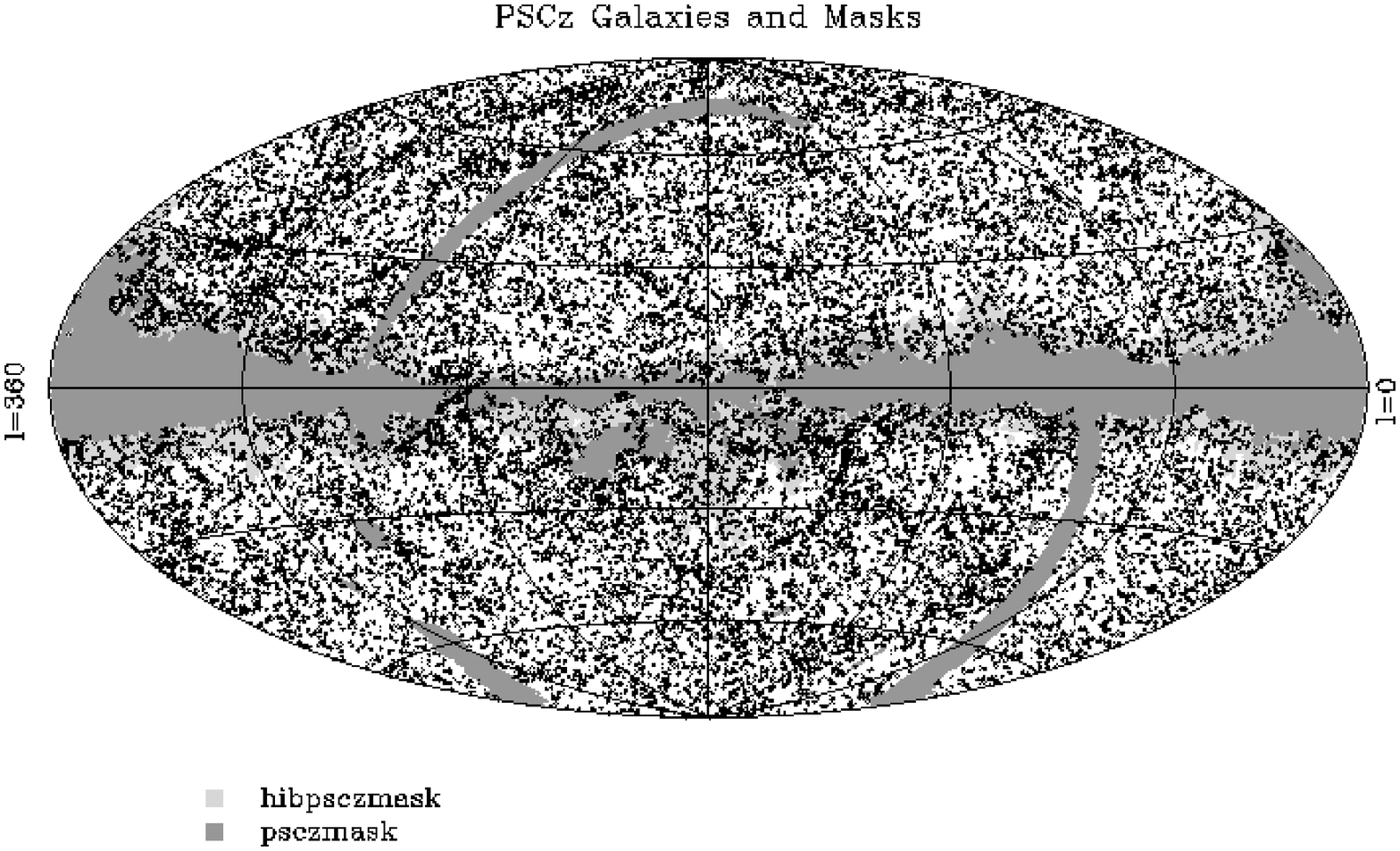,width=9cm,angle=-90}}
Figure 1. PSCz galaxy catalogue, mask and `high$|b|$' mask in galactic coordinates.
\vspace{-5pt}
\end{figure}
\begin{figure}
\centerline{\epsfig{figure=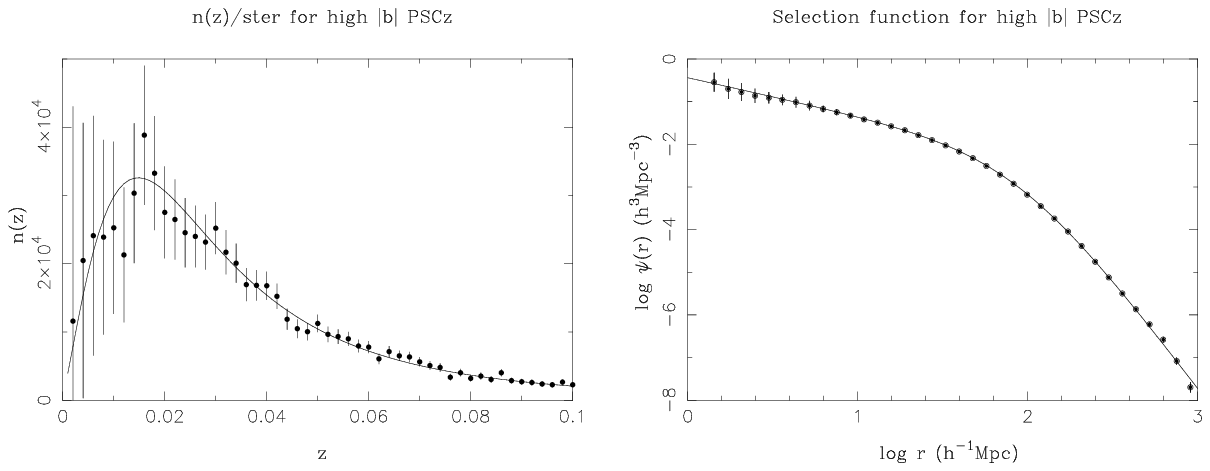,width=5cm,angle=-90}}
\vspace{5pt}
Figure 2a. $N(z)$ distribution for high latitude PSCz survey. 
Error bars are $J_3$-weighted. The line is the prediction from the selection 
function. \\ Figure 2b. 
Parametric and non-parametric selection functions for high latitude PSCz survey.
\vspace{-5pt}
\end{figure}
\begin{figure}
\centerline{\epsfig{figure=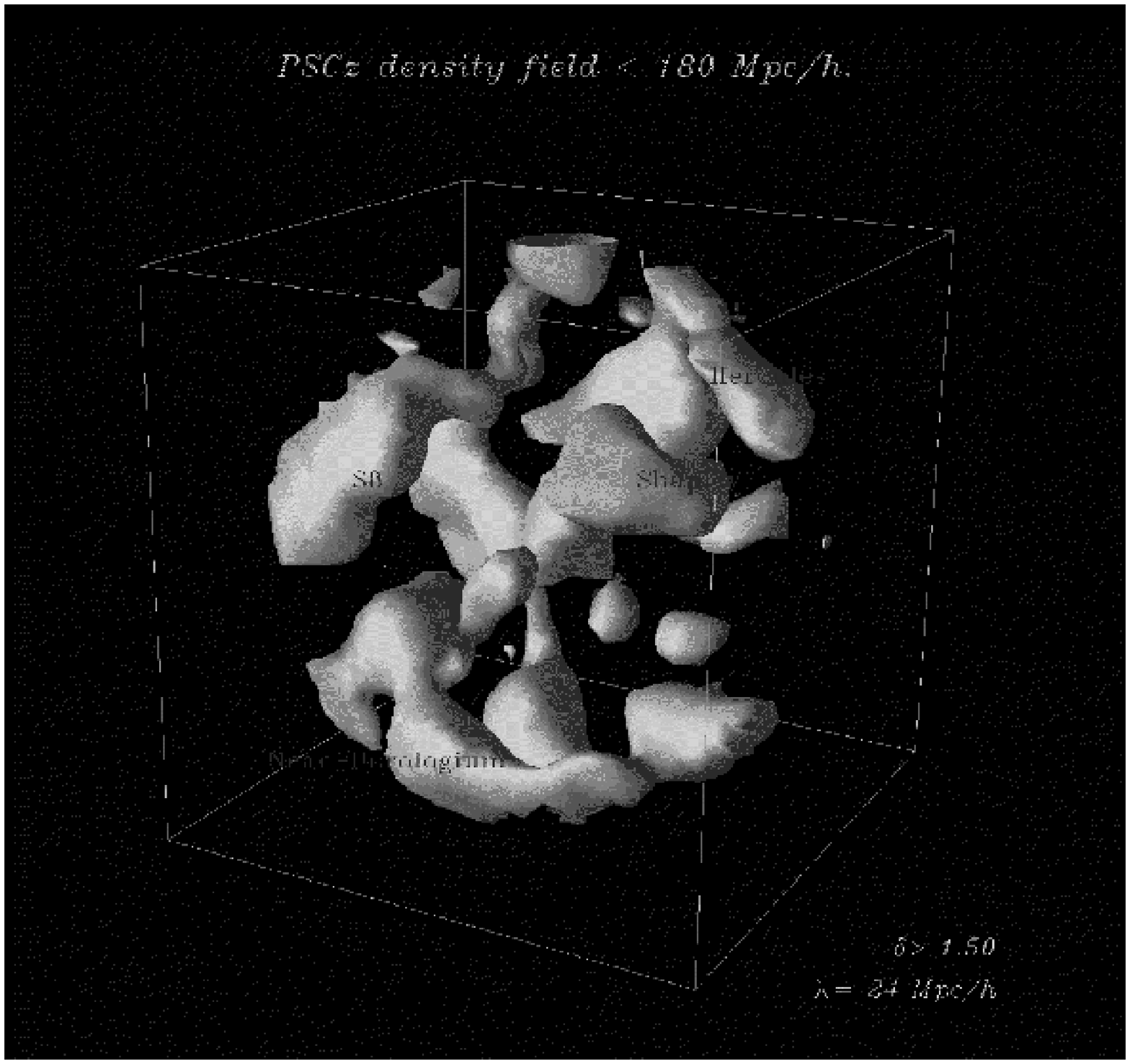,width=10cm,angle=-90}}
\vspace{5pt}
Figure 3. 3D density field of the PSCz in redshift space smoothed with a
Gaussian filter of $6 \hMpc$; 
shown is a single isodensity contour at $\delta=1.5$.
\vspace{-5pt}
\end{figure}

\begin{figure}
\vspace{-35pt}
\centerline{\epsfig{figure=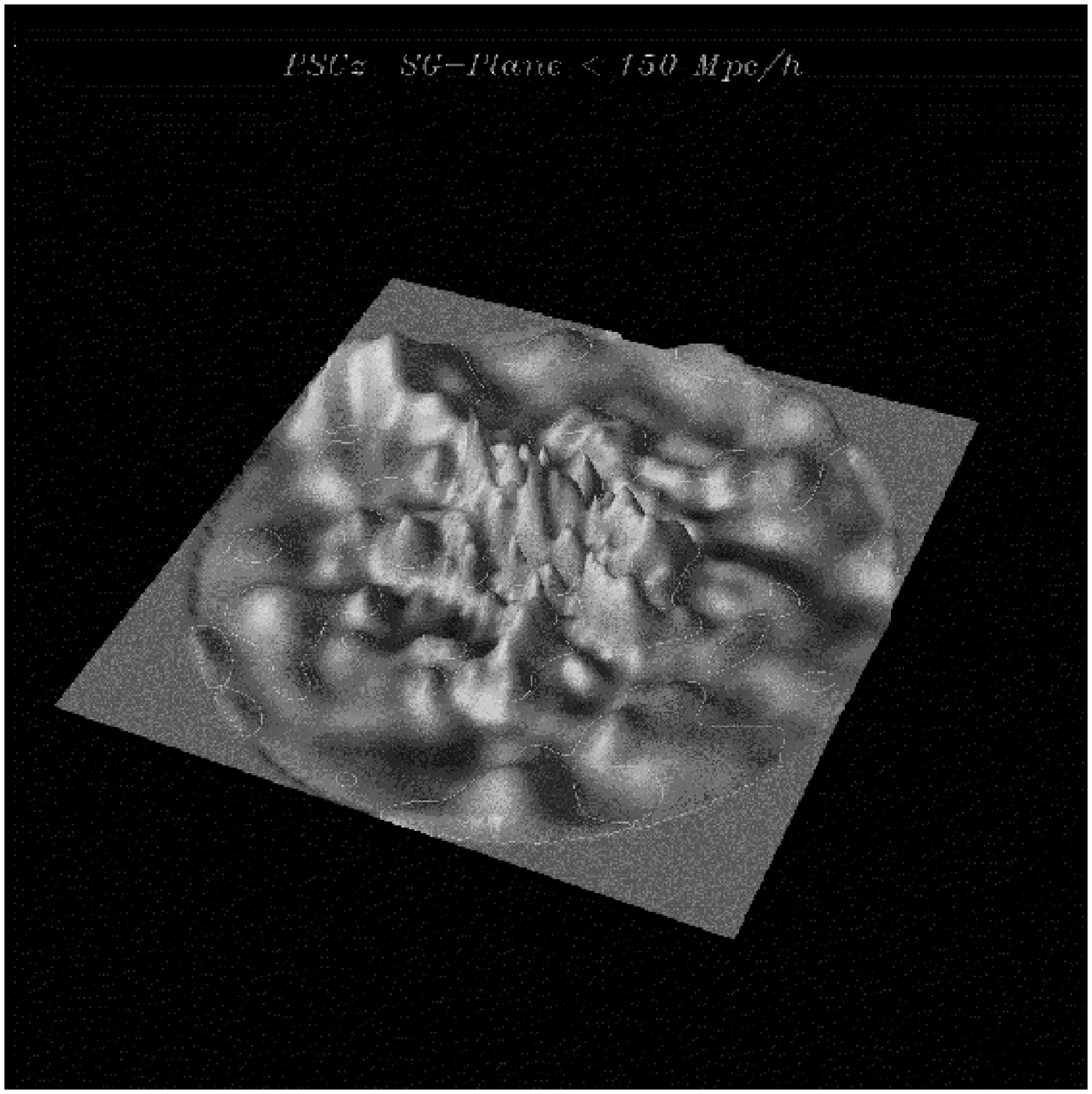,width=10cm,angle=-90}}
\vspace{5pt}
Figure 4. Real space density field of the PSCz in the Supergalactic Plane.
\vspace{-15pt}
\end{figure}

\section{A New Method for Smoothing and Interpolating Galaxy Surveys}

For dynamical tests where we wish to compare the observed distribution of galaxies 
with observed peculiar velocities, it is necessary to estimate the mass 
distribution within any masked area. Previous attempts have involved filling the 
mask with a uniform distribution, or crudely interpolating or cloning the density 
on either side. A better way was pioneered by Lahav \etal (1994), who introduced Wiener 
filtering to produce a minimum variance interpolation, 
given a prior estimate of the power spectrum and assuming linear theory. 

It is even feasible to perform such studies far into the nonlinear 
clustering regime by assuming that the field can be reasonably described
by lognormal statistics, which is indeed supported by several theoretical assessments.
In this case, fitting the log of the density field as a Fourier sum 
will lead to random phases and Gaussian Fourier amplitudes. 
The mean field for the interpolation will be the mean density. 
Our approach is to find the set of harmonics maximising the probability that 
the galaxies, assumed to be Poisson-sampled from the density field, are at 
the positions actually observed. So we have a set of galaxies at 
positions ${\bf r}_m$, and a set of basis functions ${f_n}$, 
and a set of amplitudes ${a_n}$ which we wish to determine. 
The amplitude of the underlying density field at $\bf{r}$ is 

\be
\rho({\bf r}) = \exp{\sum_n a_n f_n({\bf r})}
\ee

\noindent and the likelihood for the whole survey as a function of $\{a_n\}$ is given by

\be
\ln{{\cal L}} \{a_n\} = \ln{\prod_m \rho({\bf r}_m)} = \sum_m \sum_n a_n f_n({\bf r}_m).
\ee

\noindent The integral constraint that the total number of galaxies predicted by the 
density field equals the number actually observed is invoked via a Lagrange multiplier. This yields  $N$ equations

\be
\sum_m f_n({\bf r}_m) = \int_V f_n \psi(r) \exp{[\sum_n a_n f_n({\bf r}_m)}],
\ee

\noindent 
which states that for each harmonic the sum over its value 
at each of the galaxy positions is equal to the integral over the 
continuous density field. The equations are non-linear whenever the density field 
itself is, and we solve them by using the multidimensional Newton-Raphson technique. 
We have used a spherical harmonic expansion, and we have transformed the 
radial coordinate of the survey so as to make the selection function unity, rendering
the shot noise constant everywhere in the final maps. 
We have added a simple regularisation term to the likelihood. 
It is not mandatory for the radial and angular parts to be separable
and we have used a `tapered' mask in which the redshift out to which completeness 
is assumed is taken from the Saunders \etal (1999). 
The resulting 3-D density field is shown on the PSCz web page 
http://www-astro.physics.ox.ac.uk/$\sim$wjs/pscz.html. 
A single shell is shown in figure 5. 

\begin{figure}
\centerline{\epsfig{figure=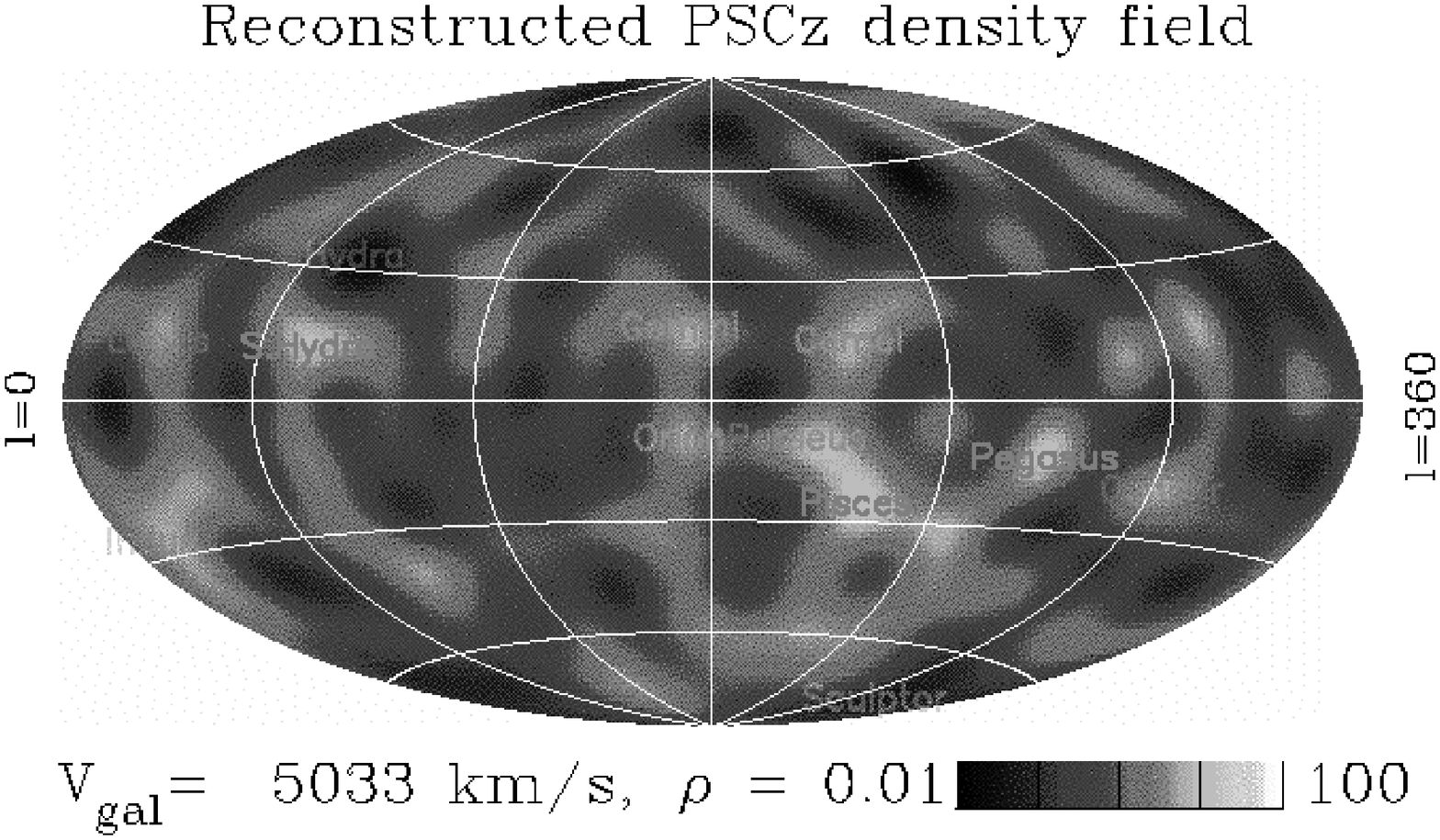,width=8cm,angle=-90}}
Figure 5. Galaxy density field at a redshift of $5000 \kms$ in galactic coordinates.
\vspace{-15pt}
\end{figure}

\section{The PSCz dipole}

The PSCz dipole has been investigated and presented by Rowan-Robinson \etal 1999
and Schmoldt \etal (1999a,b).
We here present a somewhat different analysis: we have corrected for redshift space 
distortions according to Valentine \etal (1999) and used 
the interpolation described above. 
We have weighted the gravity dipole by $4 \pi J_3 \psi(r)/(1+4 \pi J_3 \psi(r))$, 
as in Strauss \etal (1992) (and we have assumed that $4 \pi J_3=10000 h^{-3}{\rm Mpc}^3$) 
to produce a minimum-variance cumulative dipole, given our knowledge of the power 
spectrum. 
The results are shown in figures 6a and 6b. 
The suppression amounts to a factor 1.1 at $100 \hMpc$, 
2 at $180\hMpc$ and 10 at $300 \hMpc$. 
There is no evidence for any significant contribution to the dipole beyond 
150 $\hMpc$, and the angular convergence is spectacular. 
The misalignment with the CMB dipole is $20\deg$.

\begin{figure}
\centerline{\epsfig{figure=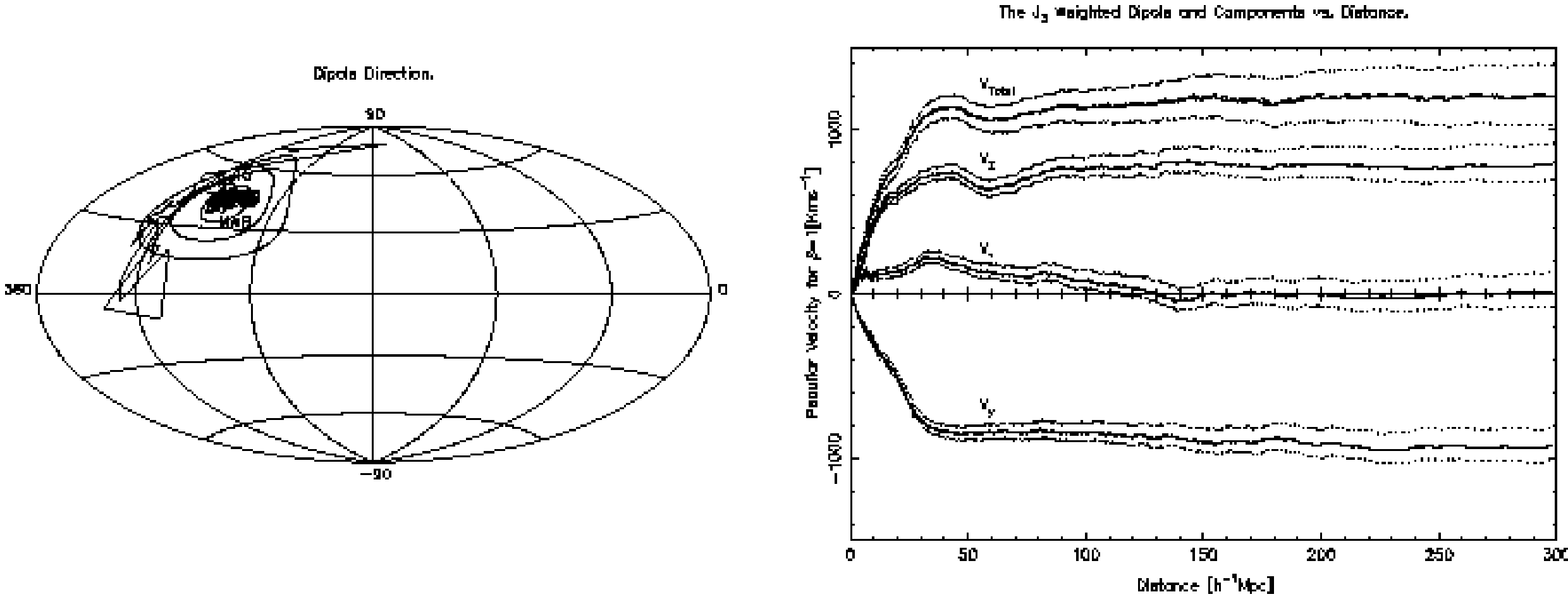,width=5cm,angle=-90}}
Figure 6a. Angular $J_3$-weighted cumulative dipole based on PIZA velocity 
reconstruction. \\
Figure 6b. The separate components and the combined magnitude. 
1 $\sigma$ shot noise error envelopes are shown.
\end{figure}

\section{Model Velocity Fields}

We have applied several different methods to obtain
a self consistent model for the density and velocity fields
from the PSCz dataset. 
All methods assume gravitational instability and linear biasing,
some are based on linear theory (Branchini \etal 1999 and Schmoldt \etal 1999b)
some others on the Zel'dovich approximation (Valentine \etal
and  D'Mellow and Taylor,
these  proceedings), while others use the Least Action Principle (Sharpe \etal 1999, Nusser \etal 1999). 
Figure 7 shows the density and velocity fields
in a slice along the Supergalactic Plane, reconstructed by 
Branchini \etal (1999).
The dominant features in the velocity fields are the infall patterns 
towards the Great Attractor (-30,20), Perseus Pisces (50,-10), 
Cetus Wall (20,-50) and Coma (5,70). The most striking property, however, 
is the large scale coherence of the velocity field, apparent as a 
long ridge along the Perseus Pisces - Virgo - Great Attractor -
Shapley Region baseline.  
The same general features are found using the other dynamical methods.

\begin{figure}
\centerline{\epsfig{figure=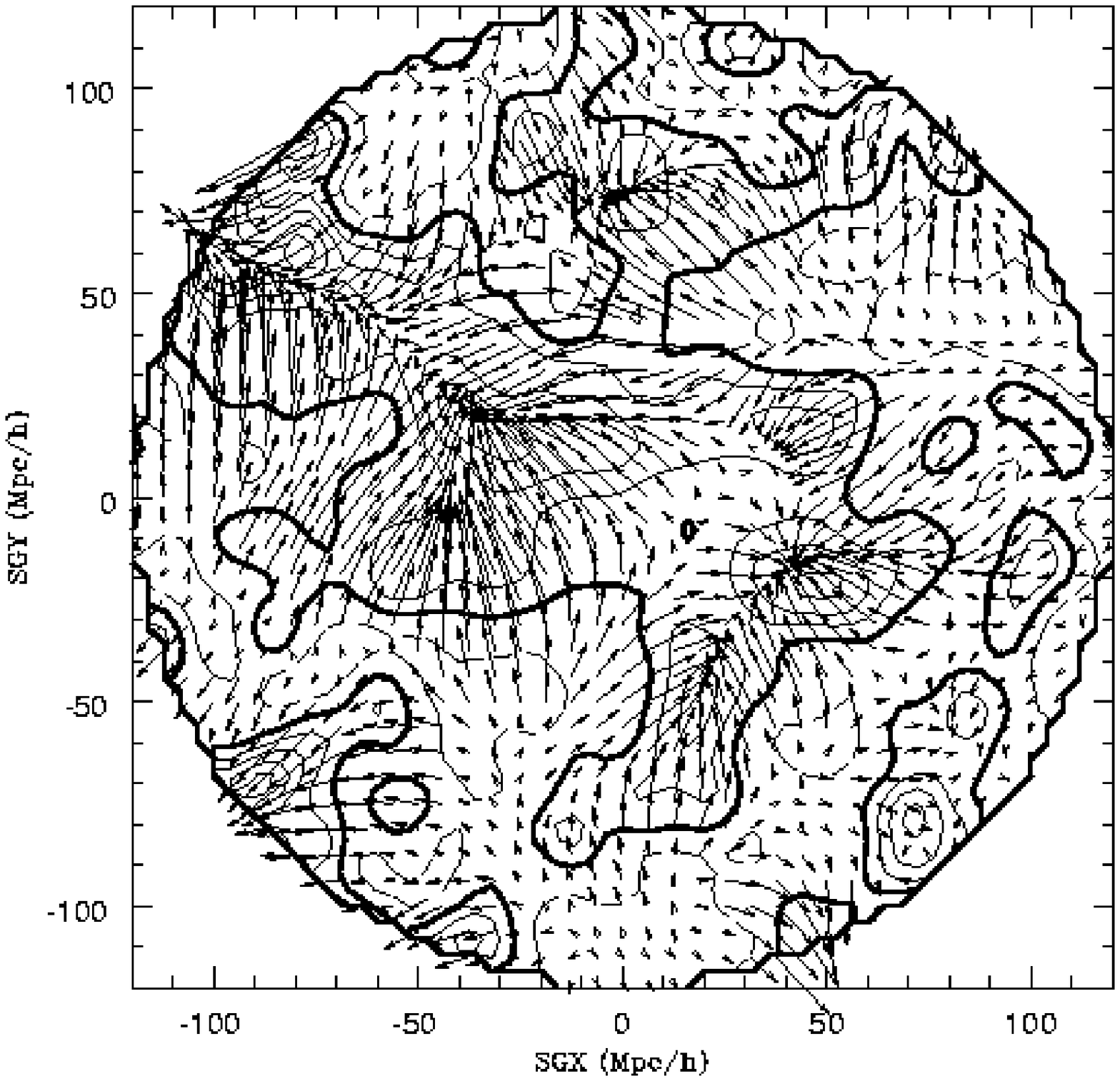,width=8cm,angle=0}}
Figure 7 Real space overdensity and velocity field smoothed with a 6 $\hMpc$
filter.  The thick line shows the $\delta=0$ contour. Positive (continuous) and 
negative (dashed) contours are plotted at steps of $\Delta \delta=0.5$
\vspace{-15pt}
\end{figure}

\section{The Value of $\beta$}

The comparison between measured peculiar velocities and 
our PSCz model gravity field allows one to measure the 
$\beta$ parameters. 
Matching the amplitude of the PSCz and CMB dipoles yields a value 
for $\beta=0.54 \pm 0.1$, consistent with  similar comparisons
by Schmoldt \etal (1999a,b) and with the likelihood analysis of
Branchini \etal (1999)  which also use 
the Mark III bulk flow, the PSCz predicted bulk flow and the local shear.
Similar values are also found by Sharpe \etal (1999) by considering
the dynamics of the PSCz galaxies in the Local Group's neighborhood, and by Tadros \etal (1999) from considering the large scale statistical distortion of the density field in redshift space.

An estimate of $\beta$ to within $10 \%$ can be achieved by
independent comparisons between observed and
predicted galaxies' velocities. An analysis along these lines  
on the basis of the observed velocities of the 
SFI catalog is in progress.

\section{Nonlinear Biasing}

The dense sampling of PSCz galaxies and the volume of the sample are 
large enough to measure the biasing relation. Narayanan \etal and Sigad
\etal (these proceedings) have presented two independent methods 
for measuring the biasing relation which they apply to the PSCz catalog.
In both cases they detect deviations from the simple linear prescriptions
which are well described by semi-analytic model for galaxy formation
(Kauffman \etal 1999, Benson \etal 1999).
Figure 8 shows the mean biasing relation for PSCz galaxies obtained by 
Sigad \etal. It is compared with semi-analytic predictions from Kauffman\etal (1999) in two different cosmological models.

\begin{figure}
\centerline{\epsfig{figure=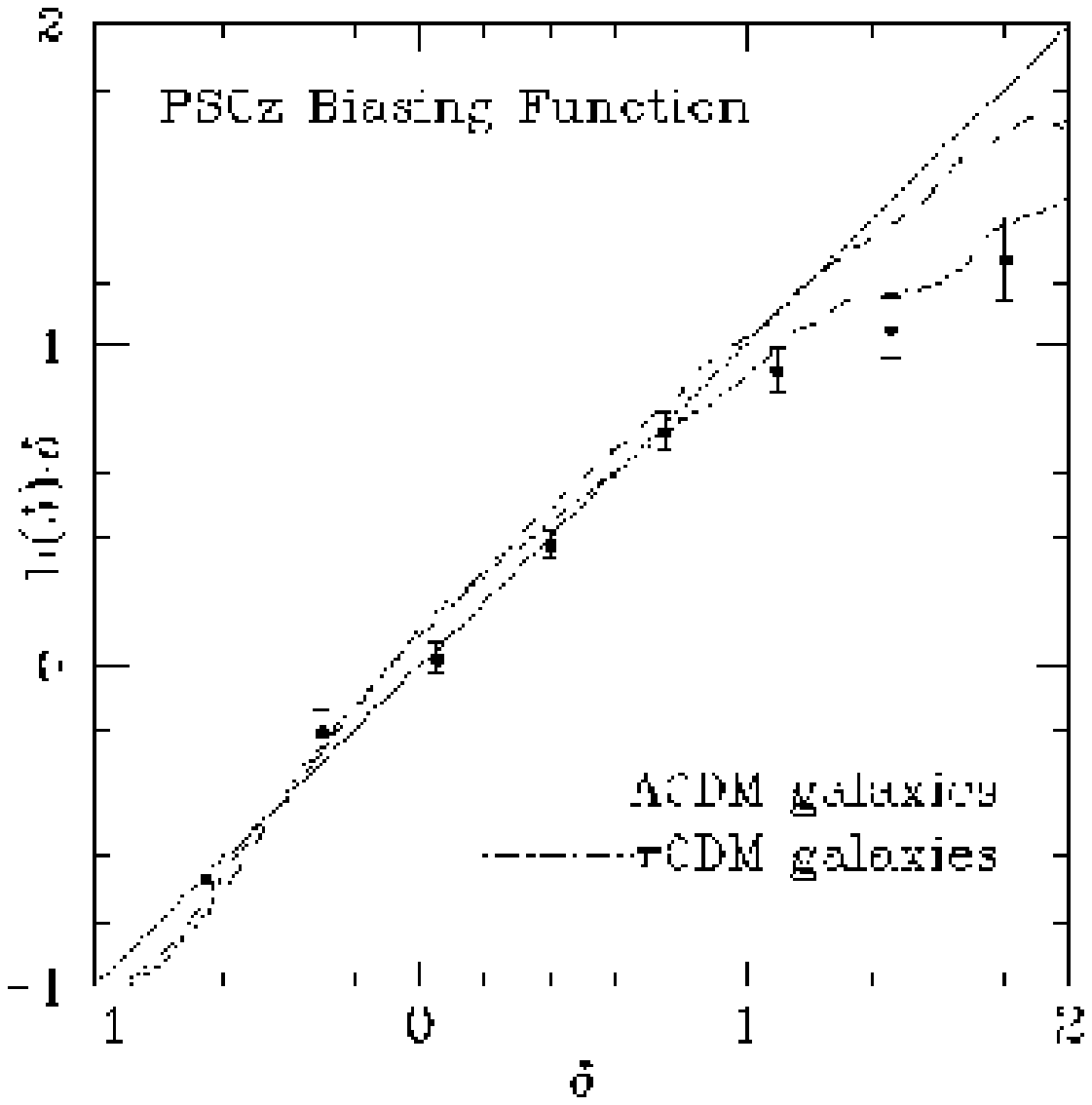,width=8cm,angle=0}}
\vspace{-2pt}
Figure 8. Biasing function for PSCz galaxies (dots). 
Errorbars represent 1$\sigma$ bootstrap errors. Predictions from semi-analytic
models are displayed for a flat CDM universe with $\Lambda=0.7$ and 
(dashed line) and for a $\tau$CDM universe (dot-dashed). 
\vspace{-15pt}
\end{figure}

\section{Acknowledgements}
The PSC-z survey has only been possible because of the generous assistance from many people in the astronomical community. We are particularly grateful to John Huchra, Tony Fairall, Karl Fisher, Michael Strauss, Marc Davis, Raj Visvanathan, Luis DaCosta, Riccardo Giovanelli, Nanyao Lu, Carmen Pantoja, Tadafumi Takata, Tim Conrow, Mike Hawkins, Delphine Hardin, Mick Bridgeland, Renee Kraan-Kortweg, Amos Yahil, Alberto Caraminana, Esperanza Carrasco, Brent Tully, and the staff at IPAC and the INT, AAT, CTIO and INOAE telescopes. We have made very extensive use of the NED, LEDA and Simbad databases.

\section{Data access}
\vspace{-5pt}
Long and short versions of the catalogue, maskfiles, notes, and an expanded version of this paper, will shortly be available via the PSCz web site \\ (http://www-astro.physics.ox.ac.uk/$\sim$wjs/pscz.html).

\end{document}